\documentclass[twocolumn,showpacs,prX]{revtex4-1}

\usepackage{ctable}
\usepackage{rotating}
\usepackage{epstopdf}
\usepackage{subfigure}
\usepackage{enumitem}
\usepackage{verbatim}
\usepackage{eqparbox}
\usepackage{graphics}
\usepackage{latexsym}
\usepackage{graphicx}
\usepackage{amsmath}
\usepackage{cleveref}

\graphicspath{{figures/}}

\begin{document}

\preprint{Amitai's paper}

\title{Detection of a Boson Star at Sgr A* through Gravitational Lensing}

\author{Amitai Y. Bin-Nun}
\email[]{binnun@gmail.com}
\altaffiliation{Department of Physics, Yeshiva University}
\affiliation{Department of Physics, Yeshiva University}

\date{\today}

\begin{abstract}
Observations of the Sgr A* region in the galactic center confirm the presence of a large amount of matter in a small volume, leading to the consensus that a black hole exists there. However, dynamical observations cannot rule out the presence of a boson star, a compact object made up of scalar particles, as both objects are far more compact than the scale of current observational constraints. While a boson star in the galactic center is disfavored for a number of theoretical considerations, we outline the first test that can directly observe a boson star. We accomplish this by studying the strong gravitational lensing of S stars resulting from the assumption of a boson star in the Galactic Center. Boson stars have an extended mass distribution  and are transparent to electromagnetic radiation, giving rise to a radial caustic curve. We calculate the brightness of images formed by stars crossing these radial caustics and show that a boson star would give rise to much brighter images than a black hole with a similar mass and that those images would be easily bright enough to be detected with upcoming instruments.
\end{abstract}

\pacs{04.40.-b, 98.62.Sb, 04.80.Cc}

\maketitle

\section{Introduction}

Over the last generation, increasingly precise observations of the Sgr A* region in the center of our galaxy have given astrophysicists exciting results \cite{genzelreview}. In particular, observations of stellar orbits in the galactic center put strong constraints on the mass contained in a small volume, leading to the conclusion that there is a black hole of about $4.31 \times  10^6$ M$_{\odot}$ and at a distance from Earth of 8.33 kpc \cite{propSgrA}.  Alternatives to a black hole have been considered, including a boson star \cite{torresetal}, a collection of scalar particles that are upheld against gravitational collapse by the Heisenberg uncertainty principle and a repulsive interaction between the component particles. This creates a compact object that comes in different sizes, but no smaller than about an order of magnitude larger than a black hole of the same mass. While there are open questions as to how a boson star could form in the galactic center and why it would not accrete baryonic matter and collapse into a black hole, there is little way to probe the distance scale that differentiates whether the mass at Sgr A* is a black hole or a boson star. \\

In this paper, we explore the possibility that the nature of this dark object could be telegraphed through observation of its secondary images. Studies have considered the properties of secondary images of stars orbiting a black hole at Sgr A*; excitingly, images created by a boson star can be much brighter because the extended mass distribution of the boson star gives rise to radial caustics. In Sec. \ref{sec:boson}, we discuss boson stars and their potential role in the galactic center. In Sec. \ref{sec:results}, we show that boson star lensing would lead to greatly enhanced secondary images. In Section \ref{sec:discussion}, we discuss the implications and prospects for these observations.

\section{Boson Stars}
\label{sec:boson}

Fundamental scalar fields have only recently been confirmed to exist in nature \cite{higgs}, but  play a key role in theoretical particle physics. A boson star (BS) is a massive collection of complex scalar particles localized by self-gravitation and self-interaction encoded in the potential $U( | \Phi^2|)$. Solutions  for the complex scalar field $\Phi$ arise from the coupled Einstein and Klein-Gordon equations:

\begin{eqnarray}
\left( \Box + \frac{ d U}{d | \Phi |^2} \right) \Phi &=& 0, \label{eq:kg1}\\
R_{\mu \nu} - \frac{1}{2} g_{\mu \nu} R &=& - \frac{8 \pi}{M^2_{\text P}}  T_{\mu \nu} \left( \Phi \right), \label{eq:einsteinboson}
\end{eqnarray}
where $\Box \equiv \left (1/ \sqrt{|g|} \right) \partial_{\mu} \left (\sqrt{|g|} g^{\mu \nu} \partial_{\nu} \right) $, $M_{\text P}$ is the Planck mass,  $T_{\mu \nu} \left( \Phi \right)$ is the stress-energy tensor derived by varying the Lagrangian of the scalar field, and units are normalized so $G=c=1$.

The system of equations (\ref{eq:kg1}-\ref{eq:einsteinboson}) is solved using a spherically symmetric ansatz for the scalar field, $\Phi \left(r,t \right)= P \left(r \right) e^{- i \omega t}$ and we assume a spherically symmetric spacetime with the metric in the area gauge:
\begin{equation}
ds^2=-A(r)dt^2+B(r)dr^2+r^2 \left(d \theta^2 + \sin^2 \theta d \phi^2 \right).
\label{eq:metric}
\end{equation}
The solution to these equations depends on the form of $U(|\Phi|^2)$. The non-interacting case $U=m^2 |\Phi|^2$, where $m$ is the mass of the component scalar particle, was studied by \cite{kaup}. They found a family of stable solutions parameterized by the central density of the field $\Phi(0)$. A potential of form $U(|\Phi|^2)= m^2 |\Phi|^2 + \lambda|\Phi|^4$ was studied by \cite{colpiboson} and other potential forms are reviewed in \cite{lieblingreview}.\\

Many superposed particles  become a boson star. As more particles are added, the boson star eventually becomes unstable against gravitational collapse, with the maximum stable mass depending on the mass of its component particles and the form of its potential \cite{torresetal}. To form a supermassive boson star of about four millions solar masses from non-interacting particles, the component particles must be incredibly tiny, of order $10^{-23}$ MeV. With a self-interaction term of order $\lambda |\Phi^4|$, with $\lambda$ set to unity, boson stars of four millions $M_{\odot}$ can be formed from particles around a MeV in mass; we concentrate on this case.

In \cite{colpiboson}, a boson star is solved in the case where self-interaction dominates. This occurs when the condition $\Lambda \equiv \frac{\lambda M^2_{\rm P}}{m^2} \gg 1$ is met. In this case, the Klein-Gordon equation is solved algebraically to leading order in $\Lambda$, yielding $P_*  = (\Omega^2/A-1)^{\frac{1}{2}}$, where $\Omega \equiv \omega/m$ and  $P_* \equiv P \Lambda^{\frac{1}{2}}$. This procedure also yields a simple set of differential equations for the metric coefficients, making it straightforward to solve the spacetime around such a boson star. The initial value $\Omega^2/A(0)$ is the central density of the boson star and determines the metric functions for the boson star spacetime. Hence, we have a single parameter family of boson stars.

\subsection{Boson Star in the Galactic Center}

While there is strong evidence for a black hole at the center of the galaxy \cite{genzelreview}, most observations are consistent with the presence of a compact boson star; it is nearly impossible to rule out such an object without direct and detailed X-ray observations at resolutions currently unavailable \cite{torresetal}. There have been discussions about potentially differentiating black holes and boson stars using accretion disk profiles \cite{guzman2010}, gravitational waves \cite{bertiGW}, and several other means \cite{schunckreview}, but no direct test is yet possible. In terms of lensing, differentiating characteristics of black holes, boson stars, and fermion stars were partially explored by \cite{dabrowskilensing, fermionlensing}. 

Existing bounds of the boson star parameter space are very weak \cite{bosonbounds}. As $\lambda$ becomes larger, the denser a boson star can be without collapsing . The densest stable boson star as $\lambda \rightarrow \infty$ occurs for $A(0) \approx 0.52$, representing a lower bound. Currently, the only upper bound on $A(0)$ in the literature comes from the constraint that the boson star's mass must lie almost entirely within the orbit of S2. Simulations we performed place this bound as $A(0) < 0.998$.  

There have been several studies using gravitational lensing of S stars to test the nature of the black hole at Sgr A* \cite{me2, me3}. Secondary images are formed by light that travels ``around'' the lens and is seen by the observer on the opposite side of the lens as the source. When light approaches close to a relativistic object the bending angle is

\begin{eqnarray}
\nonumber \alpha (x_0) &=& 2 {\int_{x_0}}^{\infty}\left(\frac{B(x)}{C(x)}\right)^{1/2}
                       \left[ \left(\frac{x}{x_0}\right)^2\frac{C(x)}{C(x_0)}\frac{A(x_0)}{A(x)}-1\right]^{-1/2} \\ 
& \times & \frac{dx}{x}- \pi,
      \label{eq:bending}
\end{eqnarray}
where $x \equiv r/2M$ and $M$ is the mass of the lens. This function is inserted into the modified Ohanian lens equation \cite{me2, ohanian} to solve for the image position
\begin{equation}
\gamma = \alpha(x_0)-\frac{D_{LS}(t)}{D_S}\theta(x_0),
\label{eq:lenseq}
\end{equation}
where $\gamma$ is the angle between the optic axis (line connecting observer and lens) and the line connecting the source to the lens, $\theta$ is the image position (to the observer, relative to the optic axis), $D_L$ is the constant distance from the observer to the lens (in this case, the distance between Earth and Sgr A*) and $D_{LS}$ is the distance, which varies over time, between the lens and the source star. A solution to this equation, $x_0$, corresponds to a unique angular position on the sky where the image appears ($\theta$). The magnification for an image is \cite{bozzamancini2009}
\begin{equation}
\mu= \frac{D_L^2}{D_{LS}^2}\frac{\sin \theta}{\frac{ d\gamma}{d \theta} \sin \gamma}.
\label{eq:mag}
\end{equation} 

Qualitatively, our argument is as follows: The turning point in the bending angle for a boson star, illustrated in Fig. \ref{fig:bending}, implies that for some value of $x_0$, $d \gamma / d \theta$ disappears and the magnification in Eq. 6 is infinite. This corresponds to the star crossing the radial caustic curve and the image appearing on the radial critical curve. Since $\alpha(x_0)$ is bounded and not monotonic, there can be two (or no) secondary images that form simultaneously. In the galactic center, these images are not be resolvable from each other with current and projected technologies, but we consider the sum of their fluxes. In the next section, we detail this calculation for several specific cases.

\section{\label{sec:results} Results}

Computing the bending angle requires determination of the metric functions. The method in \cite{colpiboson} breaks down near the point where the scalar field is predicted to be zero. In reality, the field asymptotes to zero at infinity. To generate solutions for all values of the radial coordinate, we solve the differential equation until the field is near zero. The vast majority of the star's mass is within this radius, so we smoothly sew on an exponentially decaying function onto the function $A(x)$. Using the new form of $A(x)$, we solve for $B(x)$, which results in a well-behaved function with expected properties. Finally, $A(x)$ is offset by a constant to align it asymptotically with the Schwarzschild case.  This generates smooth curves for $A(x)$ and $B(x)$ with the distinct form expected in for a boson star, serving as a toy model for studying boson star lensing.

 \begin{figure}[t]
 \begin{center}
\includegraphics[width= 0.5 \textwidth]{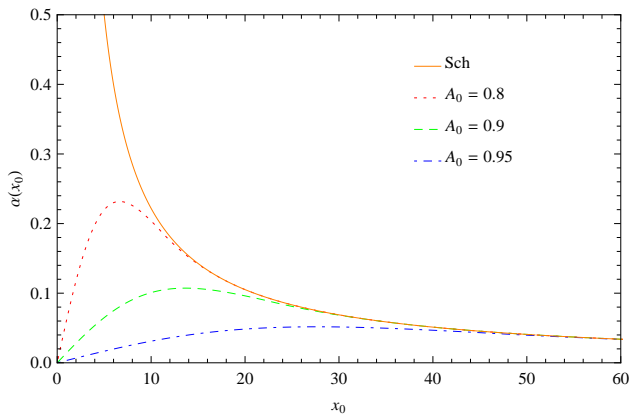}
\caption{\label{fig:bending} The bending angle curve as a function of $x_0$ for a Schwarzschild black hole and several boson stars. Boson stars are parameterized by the single quantity $A(0)$, with a smaller $A(0)$ representing a denser boson star. The denser the boson star, the sharper the peak in the bending angle function.}
 \end{center}
 \end{figure}

The functions are used to calculate the bending angle in Eq. \ref{eq:bending} as a function of $x_0$, shown in Fig. \ref{fig:bending}. 

\subsection{Boson Star Lensing}

While a black hole lens will always produce a secondary image \cite{VE2000, me}, the bending angle of a boson star is bounded, so stars will have secondary images only if they are well-aligned enough. For a particular star, an image will appear only when $\gamma$ is small enough for a real solution to Eq. 5. The largest $\gamma$ for which an image appears is approximately the maximum value of the bending angle ($\alpha_{{\rm max}}$). From Fig. \ref{fig:bending} and the properties of S stars \cite{bozzamancini2009}, it is apparent that, except for the densest boson stars, only a subset of S stars will have a secondary image and only for the portion of the stellar orbit for which $ \gamma $ is small enough.

\begin{figure}[h]
 \begin{center}
\includegraphics[width= 0.5 \textwidth]{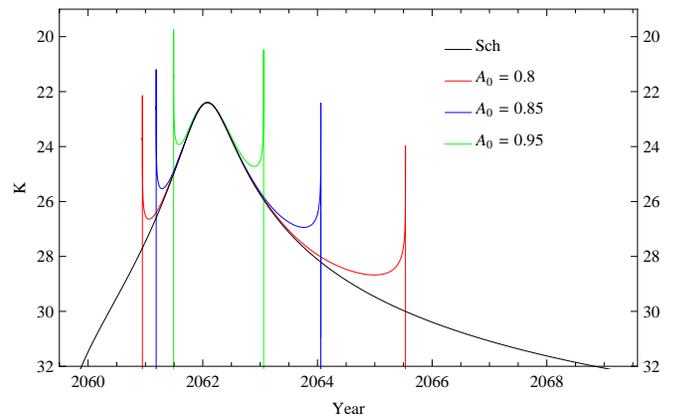}
 \caption{Apparent magnitude of the secondary image of S27 in the $K$ band if lensed by a Schwarzschild black hole or a boson star of different densities. For most of the orbital period, the secondary image for a boson star is absent, then it rapidly appears and brightens to peak brightness before fading to the same brightness as it would be in the case of a black hole (as the star moves from the caustic). Finally, the image once again rapidly brightens and then disappears as the star crosses the caustic again. As can be seen, even for the densest boson star considered, the secondary image becomes considerably brighter than in the case of a black hole. For the least dense boson star considered, the brightness peaks, for a short time, at $K = 19.7$, which is more than 10 times brighter than the peak brightness in the case of a black hole.}
 \label{fig:s27}
 \end{center}
 \end{figure}

Of particular interest is when the star crosses the caustic, which is located where $d \gamma/ d \theta =0$. The secondary image first appears when the edge of the star reaches the caustic, continues to brighten as the center of the star continues to cross the caustic, and then quickly darkens as the majority of the star crosses and moves away from the caustic. The location of the caustic depends on the parameter $A(0)$. In Fig. \ref{fig:s27}, we show the brightness over time of S27's secondary image; we chose S27 because of its bright secondary image and small minimum value for $\gamma$.

\begin{figure}[h]
 \begin{center}
\includegraphics[width= 0.5 \textwidth]{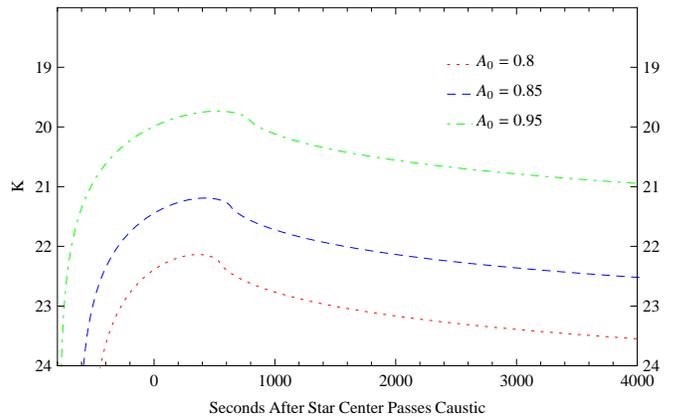}
 \caption{ The secondary image of S27 as the star crosses the caustic for three different boson star densities. The shape of all of these light curve is roughly similar --- as the boson star becomes less dense, the region of high magnification near the caustic widens and images become considerably brighter.}
 \label{fig:causticcrossing}
 \end{center}
 \end{figure}

In Fig. \ref{fig:s27}, the image seems to appear and brighten nearly instantaneously. In Fig. \ref{fig:causticcrossing}, we show the light curve as the star's center crosses the caustic. Note that for each curve in the figure, the point $t=0$ occurs at a different point in the star's orbit, corresponding to when the stellar center crosses the caustic for the corresponding density boson star.

\begin{figure}[t]
 \begin{center}
\includegraphics[width= 0.5 \textwidth]{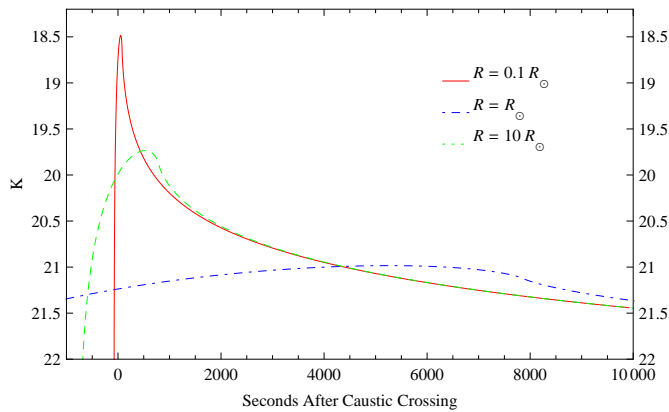}
 \caption{ The size of S27 is unknown. For three different sizes of a star with the orbit and brightness of S27, we demonstrate the resulting light curve. Smaller stellar radii result in sharper light curve peaks.}
 \label{fig:radii}
 \end{center}
 \end{figure}

For all calculations thus far, we have assumed the radius of S27 to be $R_{\odot}$. In reality, little is known about stellar characteristics in the galactic center. Stellar size is important, because larger stars do not brighten as dramatically when crossing a caustic. While S27 has not been well-analyzed, the star is likely similar to S2, which has been modeled both as an OB giant with radius approximately $10 R_{\odot}$ and as a ``stripped'' older star with a radius approximately $R_{\odot}$ \cite{s2size}. Dependence of image brightness on stellar is illustrated in Fig. \ref{fig:radii} --- we examine the case of $R = 0.1 R_{\odot}$ to illustrate the sharp magnification experienced objects such as white dwarfs, which may very well lurk invisible in the galactic center.

\begin{figure}[t]
 \begin{center}
\includegraphics[width= 0.5 \textwidth]{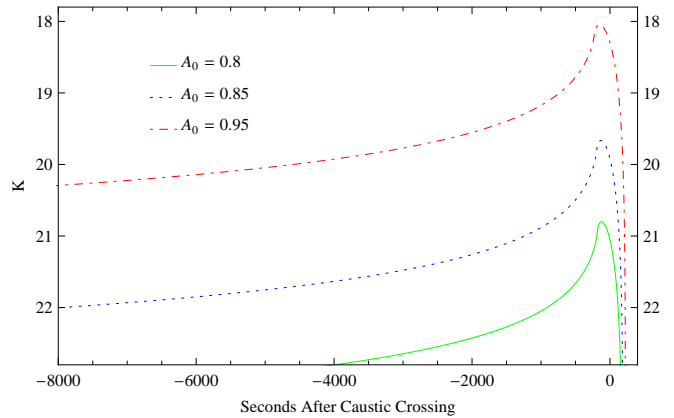}
 \caption{An analysis of S6's image as S6 crosses the radial caustic. A boson star has the ability to generate images considerably brighter than the brightest possible black hole image of $K= 20.7$.}
 \label{fig:s6}
 \end{center}
 \end{figure}

Fig. \ref{fig:s27} demonstrates the two caustic crossing events in a stellar orbit. In Fig. \ref{fig:s6}, we illustrate the light curve for the second caustic crossing of S6. For a black hole lens, the image achieves a peak brightness of $K=20.7$, which is the brightest predicted secondary image. For a boson star, the image can be brighter than $K=18$, luminous enough to be in range of next generation of instruments.

\section{Discussion}
\label{sec:discussion}

Observation of S star secondary images would be a dramatic development and would either confirm the existence of a boson star in the Galactic Center or rule out most of the parameter space. Upcoming experiments such as the Very Large Telescope Interferometer (VLTI) instrument GRAVITY or the adaptive optics camera MICADO at the European Extremely Large Telescope (E-ELT) will have the ability to observe some of these images. GRAVITY should be able to capture details as faint as $K=19$ and MICADO should be able to capture details, even in the crowded galactic center, of sources considerably fainter than $K=20$. Discussions of the relevancy of these instruments for lensing observations in the Galactic Center can be found in \cite{me2, bozza2012}.

Observing two distinct secondary images would confirm the existence of a boson star. However, the separation between the two images at their brightest is under 10 $\mu$as, well beyond the resolution of upcoming instruments. If a secondary image  was confirmed to be absent when expected for a black hole lens, that would confirm the existence of a boson star-like object. If a secondary image were to quickly brighten more than expected for a black hole lens and then fade to the black hole baseline, that would be very strong evidence of a boson star. Additionally,  images of small sources lensed by a boson star can be brighter than the primary image, so a boson star could turn invisible sources into visible secondary images. 

The photometric sensitivty of both GRAVITY and MICADO should allow lensing to be observed through astrometric measurements of centroid shift even if no image can be resolved from the quiescent state of Sgr A* ($K=17$) \cite{bozza2012}. It is important to note that a caustic crossing event due to a boson star should be easily distinguishable from the results  of Kerr black hole lensing,  as Kerr caustics are too close to the black hole on the observer's sky \cite{keetonkerr} and will not be crossed by any known S star.

The prospects for observing a lensing event may be greater than is commonly thought, as the stellar content of the Galactic Center is poorly understood \cite{ghezyouth}--- if relatively bright but small stars are abundant, we would expect a boson star to cause bright events on a frequent basis. With upcoming instruments and increased knowledge about stellar populations, gravitational lensing can serve as a valuable tool to probe distance scales that have never been previously accessible, an exciting prospect for shedding new light on this astrophysical frontier.

\section{Acknowledgements}
I thank I. Olabarrieta, B. Mundim, and M. Tierney for sharing and supporting boson star codes and R. Sheth, J. Khoury, V. Bozza, and E. Lazar for their support and helpful comments during research. Thanks are due to G. Cwilich, F. Zypman, and Yeshiva University for their hosting and  support during research for this paper.


\begin{thebibliography}{10}

\bibitem{genzelreview}
R.~{Genzel}, F.~{Eisenhauer}, and S.~{Gillessen},
\newblock Reviews of Modern Physics {\bf 82}, 3121 (2010), 1006.0064.

\bibitem{propSgrA}
S.~{Gillessen} {\em et~al.},
\newblock \apj {\bf 692}, 1075 (2009), 0810.4674.

\bibitem{torresetal}
D.~F. {Torres}, S.~{Capozziello}, and G.~{Lambiase},
\newblock \prd {\bf 62}, 104012 (2000), arXiv:astro-ph/0004064.

\bibitem{higgs}
{The CMS Collaboration},
\newblock ArXiv e-prints  (2012), 1207.7235.

\bibitem{kaup}
D.~Kaup,
\newblock Phys. Rev. {\bf 172}, 1331 (1968).

\bibitem{colpiboson}
M.~{Colpi}, S.~L. {Shapiro}, and I.~{Wasserman},
\newblock Physical Review Letters {\bf 57}, 2485 (1986).

\bibitem{lieblingreview}
S.~L. Liebling and C.~Palenzuela,
\newblock Living Reviews in Relativity {\bf 15} (2012).

\bibitem{guzman2010}
F.~S. {Guzm{\'a}n} and J.~M. {Rueda-Becerril},
\newblock \prd {\bf 80}, 084023 (2009), 1009.1250.

\bibitem{bertiGW}
E.~{Berti} and V.~{Cardoso},
\newblock International Journal of Modern Physics D {\bf 15}, 2209 (2006),
  arXiv:gr-qc/0605101.

\bibitem{schunckreview}
F.~E. {Schunck} and E.~W. {Mielke},
\newblock Classical and Quantum Gravity {\bf 20}, R301 (2003), 0801.0307.

\bibitem{dabrowskilensing}
M.~P. {Dabrowski} and F.~E. {Schunck},
\newblock \apj {\bf 535}, 316 (2000), arXiv:astro-ph/9807039.

\bibitem{fermionlensing}
N.~{Bili{\'c}}, H.~{Nikoli{\'c}}, and R.~D. {Viollier},
\newblock \apj {\bf 537}, 909 (2000), arXiv:astro-ph/9912381.

\bibitem{bosonbounds}
P.~{Amaro-Seoane}, J.~{Barranco}, A.~{Bernal}, and L.~{Rezzolla},
\newblock J. Cosml. Astropart. Phys. {\bf 11}, 2 (2010), 1009.0019.

\bibitem{me2}
A.~Y. {Bin-Nun},
\newblock \prd {\bf 82}, 064009 (2010), 1004.0379.

\bibitem{me3}
A.~Y. {Bin-Nun},
\newblock Classical and Quantum Gravity {\bf 28}, 114003 (2011), 1011.5848.

\bibitem{ohanian}
H.~Ohanian,
\newblock Am. J. Phys. {\bf 55}, 428 (1987).

\bibitem{bozzamancini2009}
V.~Bozza and L.~Mancini,
\newblock ApJ {\bf 696}, 435 (2009).

\bibitem{VE2000}
K.~S. {Virbhadra} and G.~F.~R. {Ellis},
\newblock \prd {\bf 62}, 084003 (2000), arXiv:astro-ph/9904193.

\bibitem{me}
A.~Y. Bin-Nun,
\newblock Phys. Rev. D {\bf 81}, 123011 (2010).

\bibitem{s2size}
F.~{Martins} {\em et~al.},
\newblock Astrophys. J. Lett.l {\bf 672}, L119 (2008), 0711.3344.

\bibitem{bozza2012}
V.~{Bozza} and L.~{Mancini},
\newblock \apj {\bf 753}, 56 (2012), 1204.2103.

\bibitem{keetonkerr}
A.~B. {Aazami}, C.~R. {Keeton}, and A.~O. {Petters},
\newblock Journal of Mathematical Physics {\bf 52}, 102501 (2011), 1102.4304.

\bibitem{ghezyouth}
A.~M. Ghez {\em et~al.},
\newblock The Astrophysical Journal Letters {\bf 586}, L127 (2003).

\end{thebibliography}
\end{document}